\documentclass[12pt,preprint]{aastex}

\usepackage{graphicx} 
\usepackage{epstopdf}

\def\Journal#1#2#3#4{{#4}, {#1}, {#2}, #3}
\def\ApJ{ApJ}

\def\Aph{Astropart. Phys.}

\shorttitle{S5~0716+714}
\shortauthors{Anderhub et al.}

\begin{document}


\title{Discovery of Very High Energy $\gamma$-rays from the blazar S5~0716+714}


% authors 8.7.2009  Format ApJ
%
\author{
H.~Anderhub\altaffilmark{a},
L.~A.~Antonelli\altaffilmark{b},
P.~Antoranz\altaffilmark{c},
M.~Backes\altaffilmark{d},
C.~Baixeras\altaffilmark{e},
S.~Balestra\altaffilmark{c},
J.~A.~Barrio\altaffilmark{c},
D.~Bastieri\altaffilmark{f},
J.~Becerra Gonz\'alez\altaffilmark{g},
J.~K.~Becker\altaffilmark{d}
% ...
}
\altaffiltext{a} {ETH Zurich, CH-8093 Switzerland}
\altaffiltext{b} {INAF National Institute for Astrophysics, I-00136 Rome, Italy}
\altaffiltext{c} {Universidad Complutense, E-28040 Madrid, Spain}
\altaffiltext{d} {Technische Universit\"at Dortmund, D-44221 Dortmund, Germany}
\altaffiltext{e} {Universitat Aut\`onoma de Barcelona, E-08193 Bellaterra, Spain}
\altaffiltext{f} {Universit\`a di Padova and INFN, I-35131 Padova, Italy}
\altaffiltext{g} {Inst. de Astrof\'{\i}sica de Canarias, E-38200 La Laguna, Tenerife, Spain}
% ...

\begin{abstract}
The MAGIC collaboration reports the detection of the blazar S5~0716+714
($z=0.31\pm0.08$) in very high energy gamma-rays. The observations were
performed in November 2007 and in April 2008, and were triggered by the KVA
telescope due to the high optical state of the object. 
An overall significance of the signal accounts to $S =5.8\,\sigma$ for 13.1 hours of data. 
Most of the signal ($S = 6.9\,\sigma$) comes from the April 2008 data sample during a higher optical state of the object
suggesting a possible correlation between the VHE $\gamma$-ray and optical emissions.
The differential energy spectrum of the 2008 data sample follows a power law with a photon index 
of $\Gamma = 3.45 \pm 0.54_{\mbox{stat}} \pm 0.2_{\mbox{syst}}$, and the integral flux above 400\,GeV
is at the level of $(7.5 \pm 2.2_{\mbox{stat}} \pm 2.3_{\mbox{syst}}) \times 10^{-12} \, \mbox{cm}^{-2}\, \mbox{s}^{-1}$, 
corresponding to a 9\% Crab Nebula flux. 
Modeling of the broad band spectral energy distribution indicates that 
a structured jet model appears to be more promising in describing the available data
than a simple one zone synchrotron self-Compton model.
\end{abstract}


\section{Introduction}

Blazars, a common term used for flat spectrum radio quasars (FSRQ) and
BL Lacertae objects, appear to be the most energetic types of Active Galactic
Nuclei (AGN). In these objects the dominant radiation component
originates in a relativistic jet pointed nearly towards the
observer. 
The double-peaked spectral energy distribution (SED) of blazars is
attributed to a population of relativistic electrons spiraling in the
magnetic field of the jet. The low energy peak is due to synchrotron
emission and the second, high energy 
peak is often attributed to inverse Compton scattering of
low energy photons in leptonic emission models \citep{maraschi92,dermer, bloom}. 
Models based on the acceleration of hadrons can also
sufficiently describe the observed SEDs and light curves \citep{Mannheim,
Mucke}. 
For most FSRQs and a 
large fraction of BL Lacertae objects (namely LBLs\footnote{LBL = low frequency peaking BL Lacertae}) the low energy peak is located in the energy range between
submillimeter and optical. On the other hand, for most of the sources detected
to emit VHE $\gamma$-rays (HBLs\footnote{HBL = high frequency peaking BL Lac}) the low energy peak is
located at UV to X-rays energies \citep{padovani07}. 
The high energy peak is typically at MeV--GeV energies.
Blazars are highly variable in all wavebands and the relation
between variability in different bands is a key element in
discriminating between different models.

The MAGIC Collaboration is performing Target of
Opportunity observations of sources in a high flux state in the optical
and/or X-ray band. Optically triggered observations have resulted
in the discovery of VHE $\gamma$-rays from Mrk~180
\citep{mrk180} 
and 1ES~1011+496 \citep{1011}. 
In this paper we report the results of observations of  
S5~0716+714 in November 2007 and April 2008. The observation at the latter date resulted in the discovery
of VHE $\gamma$-rays from the source as announced in \cite{ATel}.

The BL Lac object S5~0716+714 has been studied intensively at all
frequency bands. It is highly variable with rapid variations observed
from the radio to X-ray bands (Wagner et al. 1996). It has therefore
been target to several multiwavelength campaigns, the most recent one
organized by the GLAST-AGILE Support Program in July-November 2007
\citep{villata, giommi08a}.  %Villata et al. 2008, Giommi et al. 2008a. 
Due to the very bright nucleus, which outshines the host galaxy, the redshift 
of S5~0716+714 is still
uncertain. The recent photometric detection of the host galaxy \citep{nilsson} 
suggests a redshift of $z = 0.31 \pm 0.08$ which is consistent
with the redshift $z = 0.26$ determined by spectroscopy for three galaxies
close to the location of S5~0716+714 \citep{stickel}.
In the SED of S5~0716+714 the synchrotron peak is located 
in the optical band and is, therefore, classified either as
LBL \citep{nieppola}
or as IBL\footnote{IBL = intermediate frequency peaking BL Lacertae} \citep{padovani}

S5~0716+714 was detected in the MeV energy range several times at different flux levels by
the EGRET detector on board the Compton Gamma-ray Observatory
\citep{3EG}. In 2008 AGILE reported the detection of a variable $\gamma$-ray
flux with a peak flux density above the maximum reported from EGRET
\citep{AGILE}. S5~0716+714 is also on the Fermi-LAT bright source list \citep{brightsource}.
Observations at VHE $\gamma$-ray energies by HEGRA resulted in an upper limit of 
F($>1.6\,\mathrm{TeV})=3.13\times10^{-12}$ photons/cm$^2$/s \citep{Aharonian}. 
In this paper we present the first detection of VHE $\gamma$-rays from
S5~0716+714. It is the third optically triggered discovery of a VHE
$\gamma$-ray emitting blazar by MAGIC.

\section{Observations}
The MAGIC (Major Atmospheric Gamma-ray Cherenkov) telescope is a standalone imaging atmospheric Cherenkov telescope located on the Canary Island of La Palma. 
MAGIC has a standard trigger threshold of 60\,GeV for observations at low zenith angles, an angular
resolution of $ \sim 0.1^\circ$ on the event by event basis and an energy resolution
above 150\,GeV of $\sim 25\%$ \citep[see][for details]{crab}.

The Tuorla blazar monitoring program\footnote{http://users.utu.fi/kani/1m} \citep{takalo}  
monitors S5~0716+714 on a nightly basis
using the KVA 35\,cm telescope at La Palma\footnote{http://tur3.tur.iac.es/} and the Tuorla 1\,meter
telescope. At the end of October 2007 (22th, MJD=54395) the optical flux had
more than doubled (from 19\,mJy to 42\,mJy) in less than a month and, according to a predefined
criteria MAGIC was alerted.
Due to moon and weather constraints, the
MAGIC observations started 11 days later, when the optical flux had
already decreased significantly (see Fig.~\ref{fig:lc}). 
The observations were performed in the wobble mode 
\citep{daum} pointing $0.4$ deg offset from the source to enable simultaneous estimation of the background from the same field of view.
MAGIC observed the
source during 14 nights for a total of 16.8 hours. During some nights the observing conditions were rather poor and the affected data were rejected from the
analysis. The exposure time for good quality data amounts to 10.3 hours.
The zenith angle range of these observations was from 42 to 46 degrees.

In April 2008 a new bright and fast optical flare occurred.
The optical flux almost doubled within three nights (14th of April, MJD=54570: 29 mJy, 17th April:
52 mJy), and at 17th of April reached its historical maximum value. MAGIC
started the observations 5 nights later, when the moon conditions allowed.
The source was observed during 9 nights with zenith angles from 47 to 55
degrees for a total of 7.1 hours. Unfortunately, during the last 6 nights of
the observations there was strong calima wind carrying fine sand from Sahara desert 
and these data were, therefore, of bad quality and rejected from the analysis. 
The total exposure time of good quality data for this 
observation period amounts to only 2.8 hours.
The total live time of S5~0716+714 MAGIC observations in 2007 and 
2008 after data quality cuts was 13.1 hours.

\section{Data Analysis and Results}

\begin{figure}[t]
\centering
\includegraphics*[width=1.\columnwidth]{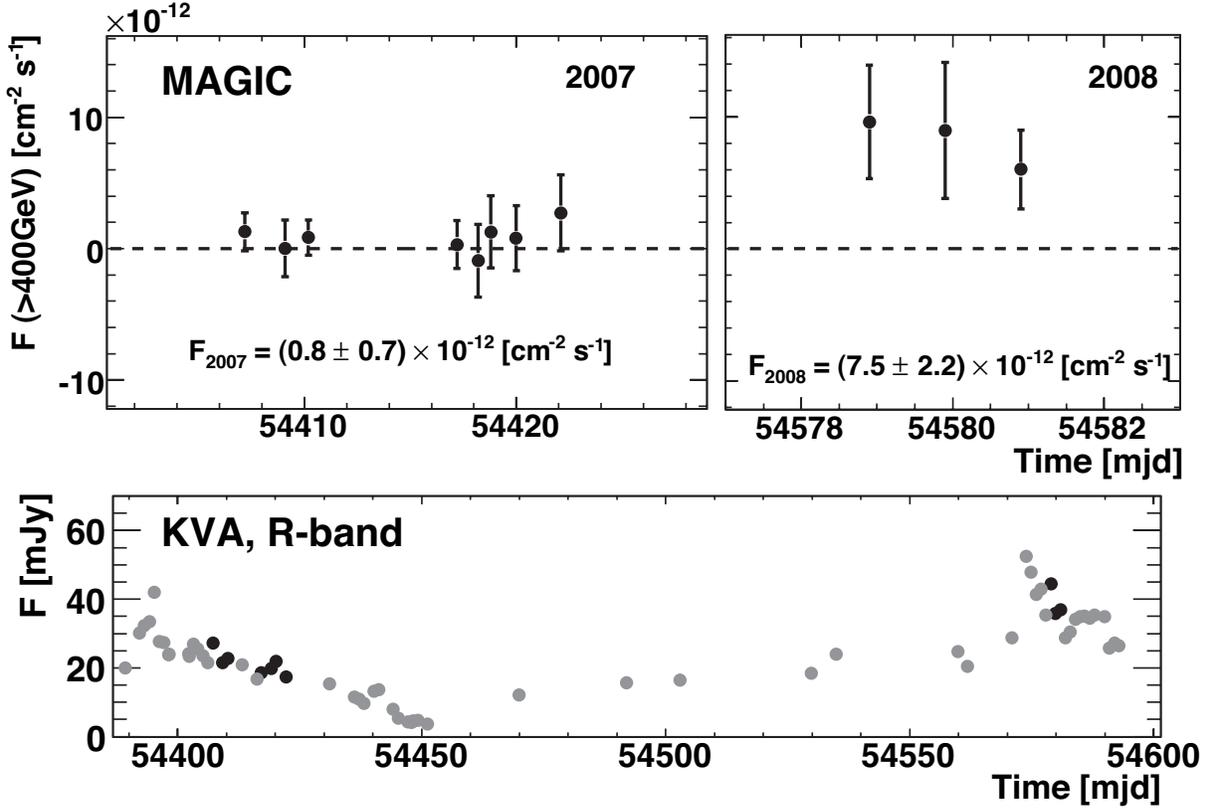}
\caption{The light curve of S5~0716+714 as measured from November 2007 until April 2008.
The day-by-day $\gamma$-ray light curve from MAGIC is shown in the upper panel for the 2007 data (left panel) and 2008 data
(right panel), whereas the optical 
KVA data are shown in the lower panel. The simultaneous optical with the MAGIC data are marked black. 
The error bars (1$\sigma$) of the optical fluxes are smaller than the points and thus not visible.}
\label{fig:lc} 
\end{figure}

\begin{figure}[t]
\centering
\includegraphics*[width=1.\columnwidth]{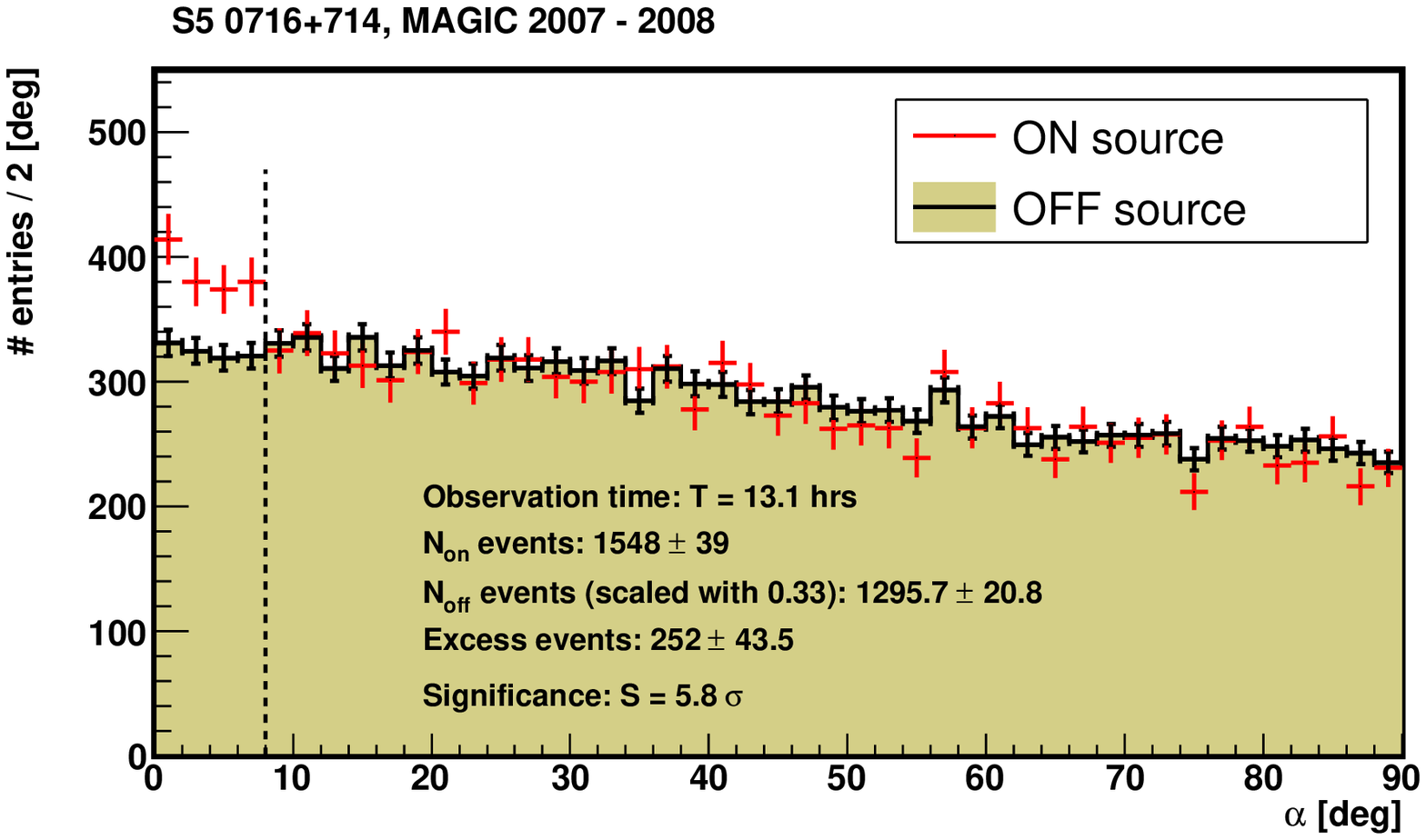}
\caption{\textsc{$\mid$Alpha$\mid$} distribution after all cuts for the total
MAGIC data sample in 2007 -- 2008. A $\gamma$-ray excess with a significance of
$5.8\,\sigma$ is found.} 
\label{fig:alpha} 
\end{figure}

The MAGIC data were analyzed using the standard analysis chain as described in 
\cite{crab,NIMA,timing}. 
In order to suppress the
%unwanted 
background showers produced by charged
cosmic rays, a multivariate classification
method known as Random Forest is used \citep{randomforest}. For every
event, the algorithm takes as input a set of image
parameters, and produces one single parameter as output,
called \textsc{Hadronness}.
The background rejection is then achieved by a cut in \textsc{Hadronness}, 
which was optimized using Crab Nebula data taken 
under comparable conditions.

The cut in \textsc{$\mid$Alpha$\mid$} that defines the signal region was also
optimized in the same way. An additional cut removed the events with a 
total charge of less than 200 photoelectrons (phe) in order to assure a better background rejection.
We used standard cuts in \textsc{Hadronness} and \textsc{$\mid$Alpha$\mid$},
which are determined to give the best significance 
for a point-like source with a flux on the 10\% level of the Crab Nebula flux.
For the given cuts and the relatively large zenith angles of the observations the
analysis threshold corresponds to 400\,GeV.
The resulting \textsc{$\mid$Alpha$\mid$} distribution after all cuts 
for the overall S5~0716+714 data sample in 2007 -- 2008 is shown in 
Fig.~\ref{fig:alpha}. An overall excess of 252  $\gamma$-like
events 
corresponding to a significance of $S = 5.8\,\sigma$ was found
(following Eq. 17 in \citet{LiMa}, 
N$_{\mathrm{on}}$ = 1548, 
N$_{\mathrm{off}}$ = 3887, alpha = 0.33). Most of the signal comes from the 2008 data sample: the analysis of
the 2008 data alone results in 176 excess events over 422 background events
corresponding to $S = 6.9\,\sigma$.  
From the 2007 data alone an excess corresponding to 
$S = 2.2\,\sigma$ was found.

\begin{figure}[t]
\centering
\includegraphics*[width=1.\columnwidth]{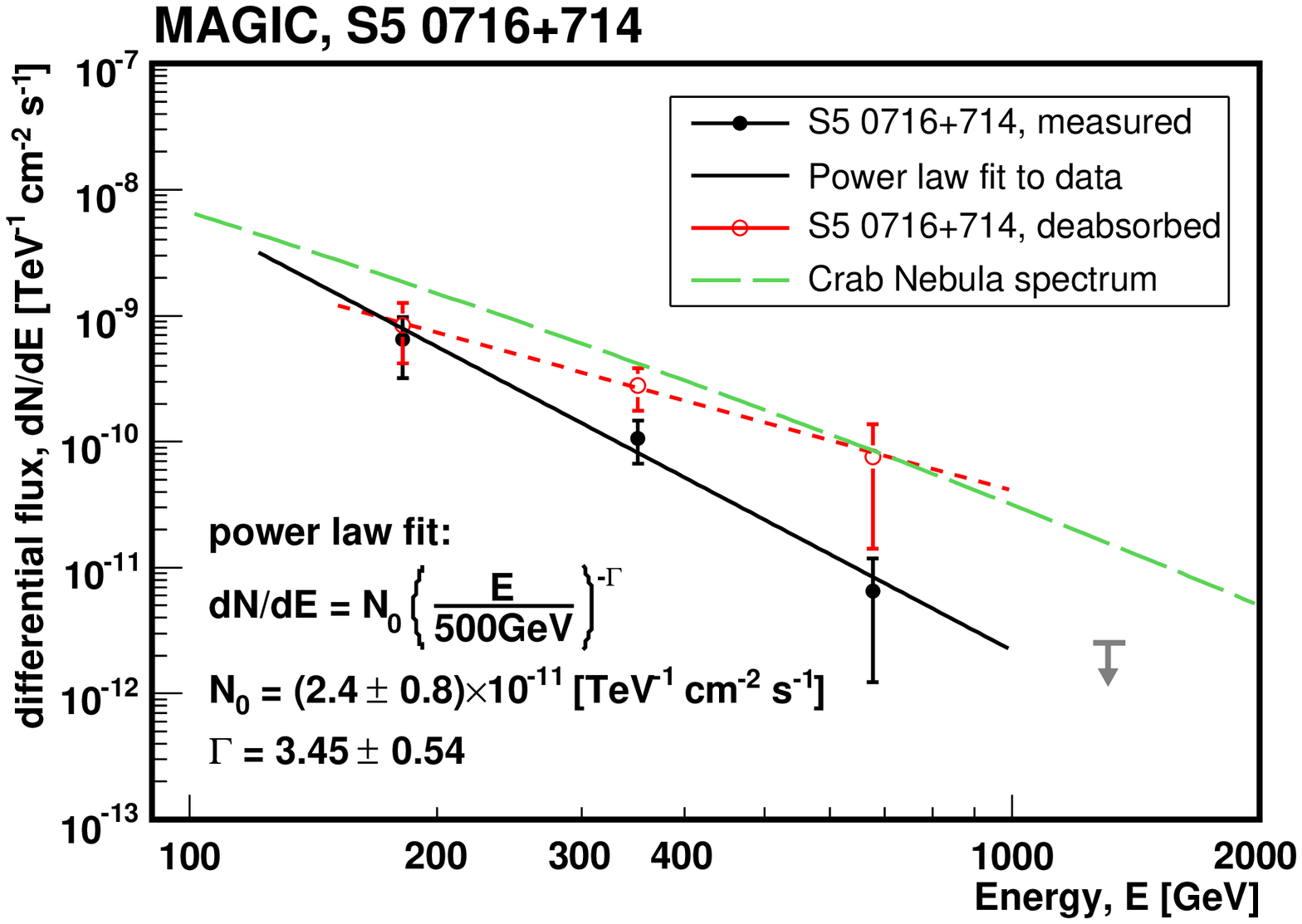}
\caption{The differential energy spectrum of S5~0716+714. Only data from April 2008 are used.
The measured (black points) as well as deabsorbed (using the EBL model
of \cite{franceschini} and assuming z=0.26, red points)
spectra are shown. 
The results from a power law fit to the measured spectrum are
shown in the plot.
The Crab Nebula spectrum is shown for comparison (green dashed line).} 
\label{fig:spec} 
\end{figure}

The day-by-day light curve as measured by MAGIC data is shown in Fig.~\ref{fig:lc} (upper
panel) together with the optical KVA light curve (lower panel).
In November 2007 the MAGIC flux above 400\,GeV is at 
$F_{\mathrm{2007}} (>0.4\mathrm{TeV}) = (0.8 \pm 0.7_{\mbox{stat}} \pm 0.2_{\mbox{syst}}) \times 10^{-11}[\mathrm{cm}^{-2} \mathrm{s}^{-1}]$,
whereas the flux is about 9 times higher in 2008:
$F_{\mathrm{2008}} (>0.4\mathrm{TeV}) = (7.5 \pm 2.2_{\mbox{stat}} \pm 2.3_{\mbox{syst}} ) \times 10^{-11}[\mathrm{cm}^{-2} \mathrm{s}^{-1}]$.
No significant variability
is seen on time scales shorter than 6 months. 
Given the limited effective exposure times these observations are not
sensitive to variability on shorter time scales. 
 The individual MAGIC points would have low significance, and an
intra-night variability by a factor of at least ten would have been required to detect it.
In the optical band,
instead, a clear variability on time scales from days to months is visible with
two distinct flares: the first in October 2007, and the second in April 2008 
(Fig.~\ref{fig:lc}, lower panel). 

The differential energy spectrum is calculated only for the April 2008 data set.
The measured and unfolded for detector effects \citep{unfolding} spectrum is shown in
Fig.~\ref{fig:spec}. The data point at $E=1.3$\,TeV has a
significance below 1\,$\sigma$ and was, therefore, converted into an
upper limit corresponding to a 95\% confidence level.  The measured spectrum
can be well fitted by a simple power law (with the differential flux
given in units of TeV$^{-1}$ cm$^{-2}$ s$^{-1}$):
\begin{equation}
\frac{\mathrm {d}N}{\mathrm{d}E\, \mathrm {d}A\, \mathrm {d}t} = (2.4\pm 0.8)\times10^{-11}(E/500\,\mathrm {GeV})^{-3.5\pm0.5} 
\end{equation}
The errors are statistical only. The systematic uncertainties are estimated to be 0.2 on the photon index and 30\%
on the absolute flux level.
Due to the energy-dependent attenuation of VHE $\gamma$-rays with low-energy
photons of the extragalactic background light \citep[EBL,][]{gould}, the VHE
$\gamma$-ray flux of distant sources is significantly suppressed.  We
calculated the deabsorbed, i.e.\ intrinsic, spectrum of S5~0716+714 
using an EBL model of  \cite{franceschini} and assuming
a redshift of $z = 0.26$. The resulting intrinsic spectrum (shown in
Fig.~\ref{fig:spec}, red points) has a fitted photon index of $\Gamma = 1.8
\pm 0.6$, which is well within the range of other extragalactic sources 
measured so far.

As the source redshift is still uncertain, we used the MAGIC spectra
to calculate upper limits to the redshift. 
We assumed two different maximum values for a possible 
hardness of the intrinsic spectrum: 
1.5, being a canonical value for a $\gamma$-ray spectrum emitted by electrons
with a spectral index of 2.0; and 0.666, being the limiting case
for  a $\gamma$-ray spectrum emitted by a monoenergetic electron distribution.
Using the method described in \citet{mazingoebel} we obtain the following upper limits
for the redshift: $z < 0.46$ 
%(assuming the hardest intrinsic index of 1.5) 
(for an assumed intrinsic spectrum with a power law photon index of 1.5)
and $z < 0.59$ 
%(assuming the hardest intrinsic index of 2/3). 
(for an assumed intrinsic spectrum with a power law photon index of 2/3).
Both limits agree with the redshift determined from the host
galaxy detection ($z = 0.31\pm0.08$) and 
from the spectroscopy of 3 nearby galaxies ($z = 0.26$).

\begin{figure}[t]
\centering
\includegraphics*[width=1.0\columnwidth,bb = 30 200 580 670]{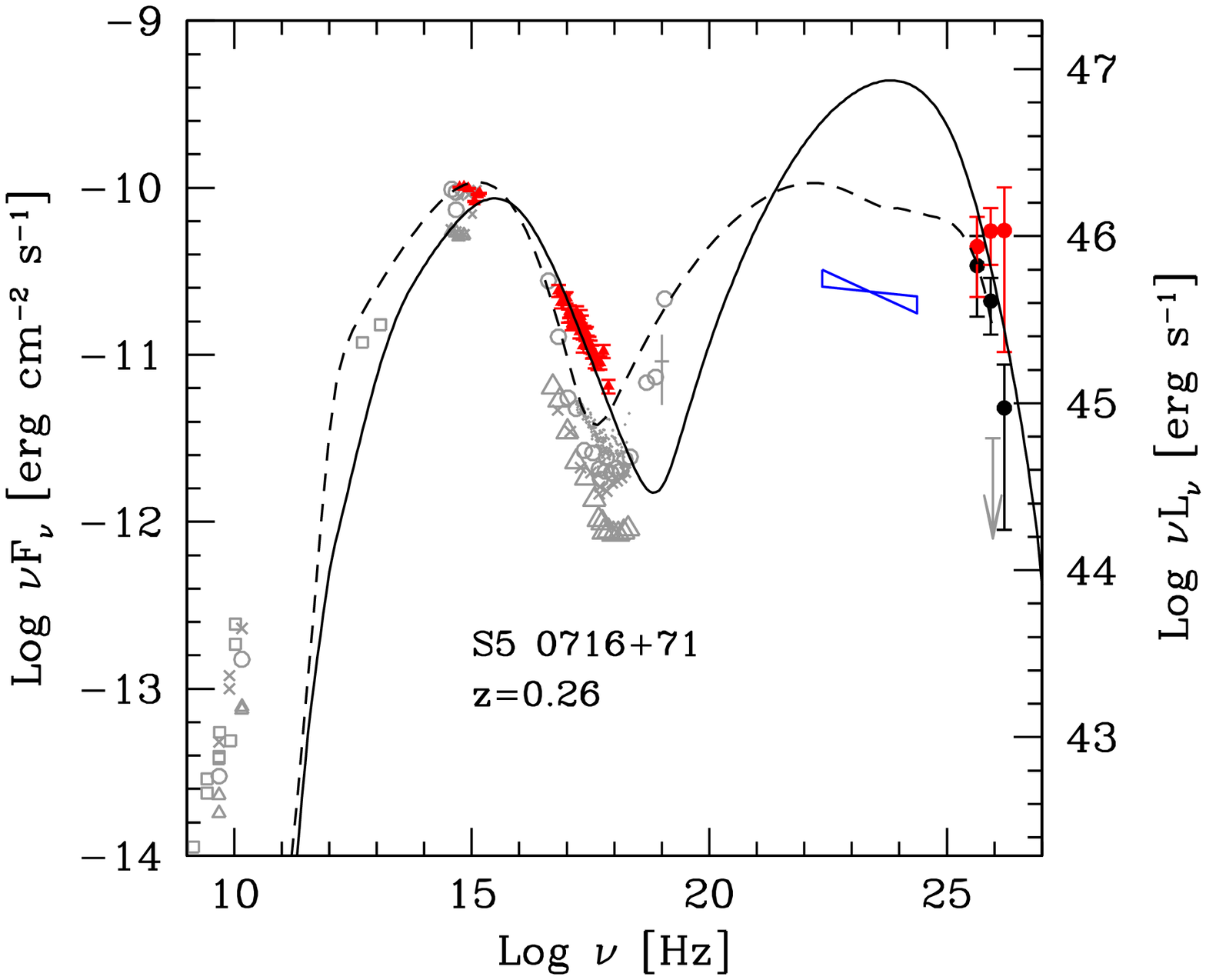}
\caption{The broad band SED of S5~0716+714. In red is the KVA (from 23 april 2008), Swift/UVOT and Swift/XRT 
(from 29 April 2008) and the deabsorped MAGIC (from 22 to 24 April 2008) data. 
The MAGIC measured flux for April 2008 is shown in black and the MAGIC upperlimit for November 2007 is shown with grey arrow.
The grey symbols show historical data \citep[see][]{tavecchio09}. The bow-tie shows the flux from the Fermi 
bright list. 
The solid line shows the overall emission calculated with a one-zone 
SSC model using the April 2008 data (red), see text for model parameters. 
The dashed line shows the emission 
calculated with the spine-layer model described in \citet{tavecchio09}.
We assume the same parameters reported there, but a 
volume three times larger.}
\label{fig:sed} 
\end{figure}

\section{Discussion}
MAGIC observed the blazar S5~0716+714 in November 2007 and April 2008, the
observations resulting in the discovery of a very high energy $\gamma$-ray
excess with a significance of $5.8\sigma$. During the November 2007
MAGIC observations the average optical flux was $\sim 20$mJy, while
in April 2008 the optical flux was $\sim 45$mJy. 
The same trend is also visible in the MAGIC data: the flux in
April 2008 is significantly higher than in November 2007. This seems
to support the indication seen in previous MAGIC observations for
other BL Lac objects \cite{mrk180,1011,bllac}, that there is
a connection between optical high states and VHE $\gamma$-ray high
states.  

In April 2008 S5~0716+714 was also in a historical high state in
X-rays \citep{giommi08b} and the optical polarization angle started to
rotate immediately after the optical maximum had been reached
\citep{larinov}.  However, the radio flux at 37 GHz did remain at a
quiescent level (A. L\"ahteenm\"aki, priv.comm). 
This energy dependent 
behavior is very similar to the one seen in BL~Lac in 2005, which was
attributed to an emission feature moving in a helical path upstream of
the VLBA core \citep{marscher}. The high VHE $\gamma$-ray flux of
S5~0716+714 observed by MAGIC might originate from such a moving
emission feature, but a more detailed study of the multiwavelength
light curves from spring 2008 are needed to confirm this.

We tried modeling the SED with a
one-zone synchrotron self Compton (SSC) model with relativistic
electrons following a smoothed broken power law energy distribution
(see \citet{tavecchio01} for a full description). 
The fit parameters are: the minimum electron energy $\gamma_{min}=10^4$,
the break energy in the electron spectrum $\gamma_b=1.9\cdot10^4$,
the maximum electrom energy $\gamma_{max}=7\cdot10^5$,
the electron spectral index below and above the break energy $n_1=2$,
$n_2=4.5$, respectively,
the magnetic field $B=0.13\,\mathrm{G}$,
the normalization factor $K=1.4\cdot10^5\, \mathrm{particles/cm}^3$,
the radius of the emitting region $R=7\cdot10^{15}\, \mathrm{cm}$ and 
the Doppler factor $\delta=30$. 
To account for the observed TeV flux corrected for intergalactic absorption the
model (continuous line in Fig.~\ref{fig:sed}) predicts a high
intensity peak in the 10\,GeV range.  Although the source was flaring
in optical at the time of the MAGIC detection, such high $\gamma$-ray
flux (more than 10 times larger than observed by EGRET and Fermi, 
open blue bow-tie in Fig.~\ref{fig:sed}) appears somewhat implausible.
Therefore, we also considered the emission from a structured jet,
modeled as a fast "spine" surrounded by a slower moving "layer"
\citep{ghisellini05}. Photons emitted by the layer are subject to IC
scattering by the relativistic electrons of the spine thus introducing
an additional contribution to the IC spectrum at energies higher than
SSC can reach. The computed model is shown as a dashed line in
Fig.~\ref{fig:sed}. The GeV excess is substantially reduced and the
spectral shape in the $\gamma$-ray domain is consistent with the
spectral indices measured at both GeV and TeV energies, though with
presumably different intensity states.

     The discovery of high flux of VHE $\gamma$-rays by MAGIC from the
     blazar S5~0716+714 in April 2008 should be compared to
     detailed light curves from other wavelengths (radio, optical and
     X-rays) as well as VLBA maps and optical polarization light curve
     in order to further investigate the origin of this interesting
     event.

\section*{Acknowledgement}

We would like to thank the Instituto de Astrofisica de 
Canarias for the excellent working conditions at the 
Observatorio del Roque de los Muchachos in La Palma. 
The support of the German BMBF and MPG, the Italian INFN 
and Spanish MICINN is gratefully acknowledged. 
This work was also supported by ETH Research Grant 
TH 34/043, by the Polish MNiSzW Grant N N203 390834, 
and by the YIP of the Helmholtz Gemeinschaft.
The authors would also like to thank the anonymous 
referee for the valuable comments and suggestions, 
which help to improve the paper.


\begin{thebibliography}{99}


\bibitem[{{Abdo} {et~al.}(2009)}]{brightsource}
%Abdo, A.\ et al.\ (the Fermi LAT Collaboration) 2009, submitted to ApJ, arxiv:0902.1559
Abdo, A.\ et al.\ (the Fermi LAT Collaboration) 2009, arxiv:0902.1559

\bibitem[{{Aharonian} {et~al.}(2004)}]{Aharonian}
Aharonian, F.\ et al.\ (the HEGRA Collaboration) 2004, A\&A, 421, 529

\bibitem[{{Albert} {et~al.}(2006)}]{mrk180}
Albert, J.\ et al.\ (the MAGIC Collaboration) 2006, ApJ, 648, L105 

\bibitem[{{Albert} {et~al.}(2007a)}]{1011}
Albert, J.\ et al.\ (the MAGIC Collaboration) 2007a, ApJ, 667, L21

\bibitem[{{Albert} {et~al.}(2007b)}]{bllac}
Albert, J.\ et al.\ (the MAGIC Collaboration) 2007b, ApJ, 666, L17

\bibitem[{{Albert} {et~al.}(2007c)}]{unfolding}
Albert, J.\ et al.\ (the MAGIC Collaboration) 2007c, Nucl. Instr. Meth. A 583, 494

\bibitem[{{Albert} {et~al.}(2008a)}]{crab}
Albert, J.\ et al.\ (the MAGIC Collaboration) 2008a, ApJ, 674, 1037

%\bibitem[{{Albert} {et~al.}(2008b)}]{3c279}
%Albert, J. et al. (the MAGIC Collaboration) 2008b, Science, 320, 1752

\bibitem[{Albert} {et~al.}(2008b)]{NIMA}
Albert, J.\ et al.\ (the MAGIC Collaboration) 2008b, Nucl. Instr. Meth. A, 594, 407 

\bibitem[{{Albert} {et~al.}(2008c)}]{randomforest}
Albert, J.\ et al.\ (the MAGIC Collaboration), 2008c, Nucl. Instr. Meth. A, 588, 424

%\bibitem[{{Albert} {et~al.}(2008d)}]{crab}
%Albert, J. et al. (the MAGIC Collaboration) 2008d, ApJ, 674, 1037

\bibitem[{Aliu} {et~al.}(2009)]{timing}
{Aliu}, E. {et~al.} (the MAGIC Collaboration), 2009, Astropart. Phys., 30, 293

\bibitem[{Bloom \& Marscher}(1996)]{bloom}
Bloom, S. D. \& Marscher, A. P. 1996, ApJ, 461, 657

%\bibitem[{{Bychova} {et~al.}(2006)}]{bychova}
%Bychova, V. S., et al. 2006, Astronomy Reports, 50, 802

\bibitem[{Chen} {et~al.}(2008)]{AGILE}
Chen, A. W.\ et al. 2008, A\&A, 489, 37

\bibitem[Daum et al.(1997)]{daum} Daum, A. {et~al.} (The HEGRA Collaboration),
\newblock \Journal{\Aph}{8}{1}{1997}.

\bibitem[Dermer \& Schlickeiser (1993)]{dermer}
Dermer, C. D. \& Schlickeiser, R. 1993, ApJ, 416, 458 

\bibitem[Franceschini {et~al.}(2008)]{franceschini}
Franceschini, A., Rodighiero, G., \& Vaccari, M. 2008, A\&A, 487, 837
%Franceschini, A. et al.\ 2008, A\&A, 487, 837

\bibitem[{{Giommi} {et~al.}(2008a)}]{giommi08a}
Giommi, P.\  et al. 2008a, A\&A, 487,49

\bibitem[{{Giommi} {et~al.}(2008b)}]{giommi08b}
Giommi, P.\ et al. 2008b, ATel 1495

\bibitem[{Ghisellini} {et~al.}(2005)]{ghisellini05} 
Ghisellini, G., Tavecchio, F., Chiaberge, M., 2005, A\&A, 
432, 401 

\bibitem[{{Gould} \& {Schr{\'e}der}(1967)}]{gould}
{{Gould}, R.~J. and {Schr{\'e}der}, G.~P.} 1967, Physical Review 155, 1408

\bibitem[{{Hartman} {et~al.}(1999)}]{3EG}
Hartman, R. C. et al.\ 1999, ApJS, 123, 79 

\bibitem[{{Larionov} {et~al.}(2008)}]{larinov}
Larionov, V. et al.\ 2008, ATel 1502

\bibitem[Li \& Ma(1983)]{LiMa} Li,T.-P., and Ma, Y.-Q.,
\newblock \Journal{\ApJ}{272}{317}{1983}.

\bibitem[{Mannheim (1993)}]{Mannheim}
Mannheim, K. 1993, A\&A, 269, 67

\bibitem[{Maraschi} {et~al.}(1992)]{maraschi92}
Maraschi, L., Ghisellini, G. \& Celotti, A. 1992, ApJ, 397, L5

\bibitem[{{Marscher} {et~al.}(2008)}]{marscher}
Marscher, A. P. et al. 2008, Nature, 452, 966

\bibitem[{{Mazin} \& {Goebel}(2007)}]{mazingoebel}
Mazin, D. \& Goebel, F. 2007, ApJ, 655, L13

\bibitem[{M\"ucke {et~al.}(2003)}]{Mucke}
M\"ucke, A.\ et al. 2003, APh, 18, 593

\bibitem[{{Nieppola} {et~al.}(2006)}]{nieppola}
Nieppola, E., Tornikoski, M., Valtaoja, E. 2006, A\&A, 445, 441

\bibitem[{{Nilsson} {et~al.}(2008)}]{nilsson}
Nilsson, K. et al. 2008, A\&A, 487, L29


\bibitem[{{Padovani} {et~al.}(1995)}]{padovani}
Padovani, P., Giommi, P. 1995, MNRAS, 277, 1477

\bibitem[{{Padovani}(2007)}]{padovani07}
Padovani, P. 2007, Ap\&SS, 309, 63


\bibitem[{{Stickel} {et~al.}(1993)}]{stickel}
Stickel, M., Fried, J. W. \& Kuehr, H. 1993, A\&AS, 98, 393

\bibitem[{{Takalo} {et~al.}(2007)}]{takalo}
Takalo, L. O. et al. 2007 ASP Series, 373, 249
%, In proceedings of ``The Central Engine of active Galactic Nuclei'' ASP Conference Series, Vol. 373, eds. L. C. Ho and J.-M. Wang

\bibitem[Tavecchio et al.(2001)]{tavecchio01} Tavecchio, F.\ et al.
\newblock \Journal{\ApJ}{554}{725}{2001}

\bibitem[Tavecchio \& Ghisellini (2009)]{tavecchio09} Tavecchio, F., Ghisellini, G. 2009, MNRAS, 394, 131

\bibitem[{{Teshima} {et~al.}(2008)}]{ATel}
Teshima, M. (for the MAGIC Collaboration) 2008a, ATel 1500

%\bibitem[{{Teshima} {et~al.} (2008)}]{ATelCrab}
%Teshima, M. (MAGIC Collaboration) 2008b, ATel 1491

\bibitem[{{Villata} {et~al.}(2008)}]{villata}
Villata, M.\ et al. 2008, A\&A, 480, 339

\bibitem[{{Wagner} {et~al.}(1996)}]{wagner}
Wagner, S. J.\ et al. 1996, AJ, 111, 2187



\end{thebibliography}
\end{document}